\documentclass[a4paper,11pt]{article}
\usepackage[usenames]{color}

\usepackage{amsfonts,amssymb}
\usepackage{theorem}
\usepackage{cite}
\usepackage{epsfig}

\newcommand{\comment}[1]{}

\begin{document}
\comment
{

\begin{center}
{\bf yang\_mills\_lattice\_prd.tex}
\end{center}
\rightline{Povo  22.10.2011, 19.00 }
\rightline{Last revised ...  Milan, 5.6.2012, 20.00 }
}
\vskip 0.4 truecm
\Large
{\bf
\centerline{On the Phase Diagram of Massive Yang-Mills}}

{

\large
\rm
\vskip 0.7 truecm

\centerline{
Ruggero~Ferrari\footnote{e-mail: {\tt ruggero.ferrari@mi.infn.it}}}

\small
\medskip
\begin{center}
Dip. di Fisica, Universit\`a degli Studi di Milano\\
and INFN, Sez. di Milano\\
via Celoria 16, I-20133 Milano, Italy\\
(IFUM-989-FT, December, 2011 )
\end{center}

\normalsize

}

\normalsize

\begin{quotation}
\rm
 {\Large Abstract:}
The phases  of a lattice gauge model for 
the massive Yang-Mills are investigated. The phase
diagram supports the recent conjecture 
on the large energy behavior of nonlinearly realized
massive gauge theories (i.e. mass {\sl \`a la }
St\"uckelberg, no Higgs mechanism), envisaging a Phase Transition 
(PT) to an 
asymptotically free massless Yang-Mills theory.

\end{quotation}
\newpage
\normalsize
\rm

\vskip 0.8 truecm

\section{Introduction}
\label{sec:int}
A novel approach to the massive Yang-Mills gauge theory has been
proposed \cite{Bettinelli:2007tq},
where the divergences are consistently removed in the loop expansion.
The removal strategy follows close the method used recently for the
nonlinear sigma model  \cite{Ferrari:2005ii}. 
It consists in the 
subtraction of the {\sl pure} pole parts of the properly normalized 
one-particle-irreducible amplitudes regularized in $D$ dimensions. 
The mathematical tools are standard, but the
proof that the technique is consistent is rather involved  
\cite{Ferrari:2005ii},
\cite{Bettinelli:2007zn},
\cite{Ferrari:2005va}.
\par\indent
Although the subtraction method is consistent with the Slavnov Taylor 
Identities,
locality and a new, {\sl ad hoc} derived, 
Local Functional Equation, the perturbative series seems to be inadequate  
for high energy processes,
thus casting some doubts on the validity of unitarity (although $SS^\dagger =1$
order by order in the loop expansion).
It has been recently suggested \cite{Ferrari:2011bx} that the cause of this 
is due to some singularities
(phase transitions) present in the parameters space 
$(\beta :=\frac{4}{g^2}, m^2)$. 
According to this scenario one can
approach the theory with the usual perturbative loop expansion 
for low-energy processes,
while the high energy processes are described by the massless Yang-Mills 
theory with
no remnants of the longitudinal polarizations. The transition between 
the two regimes may be studied by the lattice simulation. This is attempted
in the present paper, 
where we try to study the model in the 
non perturbative regime.
\par\indent
{
The present work was motivated by the arguments just outlined, but
its contents and results are valid independently from them. In fact it opens
new perspectives on the lattice gauge theories.
}
\par\indent
An intensively studied lattice gauge model
\cite{Fradkin:1978dv}-\cite{Bonati:2009pf} turns out to be the perfect
tool for the simulation of the massive Yang-Mills (i.e. mass {\sl \`a la }
St\"uckelberg). We confirm the existence of a Transition
Line (TL) which separates a {\sl confined} phase from one with physical
vector boson states. The phase LT has an end-point around 
( $\beta\sim 2.2, \,\, m^2\sim 0.381$): for smaller
$\beta$ there is a smooth transition (crossover) from one phase to the other, 
while for larger $\beta$ there are numerical indications of
singularities in the derivatives  with
respect of $m^2$ and of $\beta$  of the energy
and of the order parameter 
(the $m^2$ derivative of the free energy). The deconfined
phase is studied by using the correlators of {\sl gauge invariant}
fields. This allows a full gauge invariant  approach to the model.
We give numerical evidence of the existence of iso-vector modes
for the spin one (no spin zero is present). For the isoscalar fields there is
a faint, but persistent, signal of an energy gap both for 
spin one and zero. Far from the LT these excitations
in the iso-scalar channels are compatible with the threshold of 
two iso-vector spin one modes. However near the TL
the energy gap in the isoscalar channels is lower than the threshold, 
thus suggesting the
existence of both spin one and spin zero bound states. This effect
happens in a band attached to the TL: for large $\beta$ 
(i.e. higher than the end point value) the band is very narrow, 
while in the crossover region (low $\beta$) the onset of bound states
is smooth and on a wider region. The tantalizing question is whether
this interesting region of the phase space will ever be reached
by experiments and the presence of bound states confirmed.

\section{The Lattice Model}
\label{sec:model}

The field theory (for the $SU(2)$ group) in the continuum is
\cite{Bettinelli:2007tq}
\begin{eqnarray}
S_{\scriptstyle{YM}}
   = \frac{1}{g^2} \int d^4x \, \Big ( - \frac{1}{4} 
                              G_{a\mu\nu}[A] G^{\mu\nu}_a[A] + 
                    \frac{M^2}{2} (A_{a\mu} - F_{a\mu})^2 \Big ) \, ,
\label{stck.1}
\end{eqnarray}
where in terms of the Pauli matrices $\tau_a$
\begin{eqnarray}
A_\mu = \frac{\tau_a}{2} A_{a\mu}, \qquad
F_\mu = \frac{\tau_a}{2}F_{a\mu}:= i \Omega\partial_\mu\Omega^\dagger.
\label{stck.2}
\end{eqnarray}
$\Omega(x)$ is an element of the $SU(2)$ group, parameterized by four
real fields
\begin{eqnarray}
\Omega= \phi_0 +i \tau_a \phi_a, \qquad \Longrightarrow \phi_0^2+\vec\phi^2=1.
\label{stck.3}
\end{eqnarray}
We have
\begin{eqnarray}
F_{a\mu} = 2(\phi_0\partial_\mu\phi_a
-\partial_\mu\phi_0\phi_a + \epsilon_{abc}\partial_\mu\phi_b \phi_c).
\label{stck.4}
\end{eqnarray}
The action in eq.(\ref{stck.1}) is invariant under $g_{\scriptstyle{\scriptstyle{L}}}(x)\in SU(2)_L$ local-left 
and $g_{\scriptstyle{R}}\in SU(2)_R$ global-right transformations 
\begin{eqnarray}\!\!\!\!
\scriptstyle{SU(2)_L}\left\{
\begin{array}{l}
\Omega'(x) = g_L(x)\Omega(x) \\
A'_\mu(x) = g_L(x) A_\mu  g^\dagger_L(x)\\\,\,\,\,\,\,\,\,\,\,
 + i g_L(x) \partial_\mu
g^\dagger_L(x) 
\end{array} \right. ,\quad
\scriptstyle{SU(2)_R}\left\{
\begin{array}{l}\Omega'(x) = \Omega(x)g_R^\dagger
\\  A'_\mu (x) = A_\mu(x)
\end{array} \right.\,.
\label{stck.5}
\end{eqnarray}
{
The theory is not renormalizable due to the non-polynomial dependence
on $\vec\phi$ explicit in eq. (\ref{stck.4}). We refer to the Refs.
\cite{Bettinelli:2007tq} and  \cite{Ferrari:2005ii}
on the new strategy suggested for the consistent subtraction
of all the ultraviolet divergences. 
}
\par
The lattice model is constructed by assuming a nearest neighbor interaction 
and by requiring a na{\"\i}ve mapping into the action (\ref{stck.1}) 
in the limit of zero lattice spacing. 
{It is very important to construct a model invariant under the
discretized versions of eqs. (\ref{stck.5}).
}
The link variable is taken to be
\begin{eqnarray}
U(x,\mu)\simeq \exp(-i a A_\mu(x)).
\label{lat.1}
\end{eqnarray}
{
Both $U(x,\mu)$ and the site variable $\Omega(x)$ are elements
of $SU(2)$.
}
Thus the action is ($\beta = \frac{4}{g^2}$)
\begin{eqnarray}
S_E=\!\! \frac{\beta}{2} \,\, {\mathfrak Re}\sum_\Box Tr(1-  U_\Box)
{
+ \frac{\beta}{2} M^2a^2 {\mathfrak Re}
\sum_{x\mu}Tr\Bigl\{1- \Omega(x)^\dagger U(x,\mu)\Omega(x+\mu)\Bigr\}
}
,
\label{lat.2}
\end{eqnarray}
where the sum over the plaquette is the Wilson action \cite{Wilson:1974sk}
and the mass term has the (Euclidean) continuum limit
\begin{eqnarray}&&
\frac{\beta}{2} M^2a^2 {\mathfrak Re}
\sum_{x\mu}Tr\Bigl\{1- \Omega(x)^\dagger U(x,\mu)\Omega(x+\mu)\Bigr\}
\nonumber\\&&
\to
\frac{M^2}{g^2} \int d^4x Tr \,\, \Big\{(A_\mu - i\Omega\partial_\mu\Omega^\dagger)^2
\Big\}.
\label{lat.3}
\end{eqnarray}
In the simulation the $Tr\{1\}$ is omitted. Thus the action becomes
\begin{eqnarray}\!\!\!\!\!\!
S_E\to - \frac{\beta}{2} \,\, {\mathfrak Re}\sum_\Box Tr( U_\Box)
- \frac{\beta}{2} m^2 {\mathfrak Re}
\sum_{x\mu}Tr\Bigl\{ \Omega(x)^\dagger U(x,\mu)\Omega(x+\mu)\Bigr\}
\label{lat.2.2}
\end{eqnarray}
From now on the dimensionless parameters are $\beta$ and $m^2$. 
We work in $D=4$,
however the symbol $D$ is kept in some equations.
In the paper we will consider also the model with $m^2 \to - m^2$. 
\section{Simulation}
\label{sec:sim}
The partition function is obtained by summing over all configurations given
by the link variables and the gauge field $\Omega$
\begin{eqnarray}
Z[\beta, m^2 , N] = \sum_{\{U,\Omega\}} e^{-S_E},
\label{sim.1}
\end{eqnarray}
where $N$ is the number of sites. \par
In principle the integration over $\Omega(x)$ is redundant, 
since by a change of variables
($U_\Omega(x,\mu):=\Omega(x)^\dagger U(x,\mu)\Omega(x+\mu))$ 
we can factor out the volume of the group. $Z[\beta, m^2,N]$ becomes
\begin{eqnarray}
\!\!\!\!\!\!
 \Big[\sum_{\{\Omega\}}\Big]\sum_{\{U\}} \exp \beta\Big(\frac{1}{2} \,\, {\mathfrak Re}\sum_\Box Tr\{ U_\Box\}
+ \frac{1}{2} m^2 {\mathfrak Re}
\sum_{x\mu}Tr\{ U(x,\mu)\}
\Big).
\label{sim.2}
\end{eqnarray}
In eq. (\ref{sim.2}) the integration over $\Omega$ has disappeared;
consequently $\Omega$ in eq. (\ref{sim.1}) does not describe any 
degree of freedom. In that respect we are at variance with other
approaches to the same action (\ref{lat.2}) as in 
\cite{Fradkin:1978dv}-\cite{Bonati:2009pf}, 
where the field $\Omega$ is thought of as a Higgs
field with frozen length.
In eq. (\ref{sim.1}) we force the integration over the
gauge orbit $U_\Omega$  by means of the explicit sum over  $\Omega$.
In doing this we gain an interesting theoretical setup of the model;
in practice, our formalism is fully gauge invariant (Section \ref{sec:vec}).
Moreover by forcing the integration over the gauge orbit $U_\Omega$
we get  results which are  less noisy than those obtained by using only
the integration over the link variables in (\ref{sim.2}). 
%
%
%
%
%
\section{Functionals and Order Parameter}
\label{sec:fop}
In this model we can study the energy-per-site functional
\begin{eqnarray}&&
\!\!\!\!\!\!\!\!\! E =  \frac{1}{ N }
\frac{\partial}{\partial \beta}  \ln Z
\nonumber\\&&\!\!\!\!\!\!\!\!\!
= \frac{1}{2 N }\Big\langle {\mathfrak Re}\sum_\Box 
  Tr\{U_\Box \} 
+ m^2\sum_{x\mu} Tr\{\Omega^\dagger(x)U(x,\mu)\Omega(x+\mu) \} \Big \rangle 
,
\label{fop.1}
\end{eqnarray}
where $\langle ~ \rangle$ denotes the mean value by the
Boltzmann weight of eq. (\ref{sim.1}).
{
Moreover we introduce the {\sl magnetization}, i.e. the response
to the applied $m^2$ 
\begin{eqnarray}
{\mathfrak C}=\frac{1}{D N\beta } \frac{\partial}{\partial m^2} 
\ln Z
=  \frac{1}{2 ND }\Big\langle {\mathfrak Re}
 \sum_{x\mu} Tr\{\Omega^\dagger(x)U(x,\mu)\Omega(x+\mu) \} \Big \rangle 
.
\label{fop.2}
\end{eqnarray}
}
Then we have the plaquette energy
\begin{eqnarray}
E_P= \frac{2}{D(D-1)N}\Bigg\langle \frac{1}{2  }{\mathfrak Re}\sum_\Box 
  Tr\{U_\Box \}  \Bigg \rangle
=  \frac{2}{D(D-1)}\Big[E - Dm^2{\mathfrak C}\Big]
.
\label{fop.3}
\end{eqnarray}
There are some simple properties that will be of some help
in the sequel. Under the mapping
\begin{eqnarray}
U(x,\mu) \to - \,\, U(x,\mu)
\label{fop.4}
\end{eqnarray}
the Wilson action is invariant while the mass part
changes sign. The measure of the group integration is
invariant, then we have from eqs. (\ref{sim.1}), (\ref{fop.1}) and (\ref{fop.2})
\begin{eqnarray}&&
Z[\beta,-m^2,N]= Z[\beta,m^2,N]
\nonumber\\&&
E[\beta,-m^2,N] = E[\beta,m^2,N]
\nonumber\\&&
{\mathfrak C}[\beta,-m^2,N] =  -\,\,{\mathfrak C}[\beta,m^2,N].
\label{fop.5}
\end{eqnarray}
{
We argue that ${\mathfrak C}$ is a valid order parameter. This
will be shown in the next Sections. We briefly recollect the main
points.
Its suscpetibility ($\partial{\mathfrak C}/\partial m^2$)
has a cusp-like behavior in $m^2$ on the TL for $\beta$ greater than
the end point value. It is odd under the change of sign of 
$m^2$. Moreover the implementation of the global $SU(2)_R$ 
symmetry  is drastically
different in the far away regions where   ${\mathfrak C}\sim 0$
and  ${\mathfrak C}\pm 1$. In the first region  $SU(2)_R$ charged
fields are screened or confined, while in the second deconfined modes
are present. 
}

%
%
%
\section{The Vector Meson Fields}
\label{sec:vec}
Our approach allows the presence of  $SU(2)_L$ \underline{gauge}
\underline{invariant}  fields. Let us consider
\begin{eqnarray}
C(x,\mu) := \Omega^\dagger(x) U(x,\mu)\Omega(x+\mu)= C_0(x ,\mu) + i\tau_a C_a(x,\mu),
\label{vec.1}
\end{eqnarray}
which, according to eqs.(\ref{stck.5}), are invariant under 
local $SU(2)_L$ transformations.
$C_0(x,\mu)$ is the mass term density in the action (\ref{lat.2.2}) and it is a 
$SU(2)_R$ - scalar (isoscalar),
while $C_a(x,\mu)$ are vectors under the same group of
transformations (isovectors). 
Since $C(x,\mu)\in 
SU(2)$, we get that all fields are real and constrained by
\begin{eqnarray}
C_0(x ,\mu)^2 + \sum_a C_a(x,\mu)^2 =1.
\label{vec.1.1}
\end{eqnarray}
Moreover we expect the vacuum to be invariant
under $SU(2)_R$ global transformations (\ref{stck.5}) and therefore
\begin{eqnarray}
\langle C_a(x,\mu)\rangle =0, \qquad a=1,2,3, \qquad \forall (x,\mu).
\label{vec.2.0}
\end{eqnarray}

\noindent
The order parameter (\ref{fop.2})
\begin{eqnarray}
{\mathfrak C}= \frac{1}{D N}\sum_{x\mu}\Big\langle
C_0(x,\mu) 
\Big\rangle
\label{vec.2}
\end{eqnarray}
is the conjugate of the mass parameter $m^2$. 
{
According to eq. (\ref{fop.5}) we expect that
\begin{eqnarray}
\lim_{m^2=0}{\mathfrak C}= 0, \qquad
\lim_{m^2\to\infty}{\mathfrak C}= 1.
\label{vec.22.4}
\end{eqnarray}
}
\par\noindent
Beside the order parameter, it is important to study the following 
connected correlators.
They will provide the essential characterization of the deconfined 
phase of the system.
\begin{eqnarray}&&
C_{ab,\mu\nu}(x,y):=
\Big\langle C_a(x,\mu)C_b(y,\nu)
\Big\rangle_C
\nonumber\\&&
C_{0b,\mu\nu}(x,y) :=
\Big\langle C_0(x,\mu)C_b(y,\nu)
\Big\rangle_C
\nonumber\\&&
C_{00,\mu\nu}(x,y) :=
\Big\langle C_0(x,\mu)C_0(y,\nu)
\Big\rangle_C.
\label{vec.2.1}
\end{eqnarray}
In order to investigate the transition between phases we consider also
the susceptibility
\begin{eqnarray}&&
 \frac{\partial}{\partial m^2}{\mathfrak C}=  
\frac{\beta}{D N}\sum_{x\mu}\sum_{y\nu}\Big\langle
C_0(x,\mu) C_0(y,\nu) \Big\rangle_C
\nonumber\\&&
= 
\frac{\beta}{D N}\sum_{x\mu}\sum_{y\nu}\Big(\Big\langle
C_0(x,\mu) C_0(y,\nu) \Big\rangle
-\Big\langle C_0(x,\mu)\Big\rangle \Big\langle C_0(y,\nu) \Big\rangle\Big).
\label{vec.2.3}
\end{eqnarray}
It should be noticed that the mean square error of ${\mathfrak C}$ is
related to its derivative
\begin{eqnarray}&&
\frac{\partial}{\partial m^2}{\mathfrak C}
= \beta D N \Big\langle\Big ({\mathfrak C}-\langle{\mathfrak C}\rangle \Big)^2
\Big\rangle.
\label{vec.2.4}
\end{eqnarray}
This relation is very important for numerical simulations. 
If the derivative of ${\mathfrak C}$ had a finite limit for $N\to \infty$, the 
standard deviation would have a $1/\sqrt N$ behavior. 
If instead the derivative
diverges then the standard error might not have a decreasing 
behavior by increasing the lattice size $N$. If this is the case,
then the calculation of the derivative by using the heat bath 
yields a noisy signal. The noise might not decrease by increasing the 
lattice size, as expected in the normal case.
%
\subsection{The $SU(2)$ Right Symmetry}
If the $SU(2)_R$  symmetry is unitarily implemented then we expect
\begin{eqnarray}&&
C_{ab,\mu\nu}(x,y)=0, \qquad {\rm if} \quad a\not =b,  
\quad a,b=1,2,3
\nonumber\\&&
C_{0b,\mu\nu}(x,y)= 0 .
\label{vec.3}
\end{eqnarray}
The energy gap in the correlator in $C_{00,\mu\nu}(x,y)$ might 
set on above the two-particle threshold. However there is an
interesting possibility that the gap (both for spin one and spin zero)
shows up below this threshold, thus suggesting the existence
of bound states.
%

\subsection{The Continuum Limit of $C$}
By a similar argument used in Section \ref{sec:model}
we study the continuum limit of $C(x,\mu)$.
We have
\begin{eqnarray}
C(x,\mu)= \Omega^\dagger(x)(1-ia
A_\mu(x))(\Omega(x)+a\partial_\mu\Omega) + {\cal O}(a^2).
\label{vec.5}
\end{eqnarray}
Thus for $C_1,C_2,C_3$ one gets
\begin{eqnarray}
i\tau_aC_a(x,\mu) 
= -ia \Omega^\dagger \Big(A_\mu(x) 
- i \Omega\partial_\mu \Omega^\dagger \Big)\Omega  + {\cal O}(a^2).
\label{vec.5.1}
\end{eqnarray}
While for $C_0$ we can use the result of Section \ref{sec:model}, eqs.
(\ref{lat.2}) and (\ref{lat.3})
\begin{eqnarray}&&
C_0(x,\mu) =  1 - \frac{a^2}{4}Tr \,\, \Big\{(A_\mu - i\Omega\partial_\mu\Omega^\dagger)^2
\Big\}+ {\cal O}(a^4). 
\label{vec.6}
\end{eqnarray}
Notice that the dominant terms in eqs. (\ref{vec.5.1}) and (\ref{vec.6})
are $SU(2)$ local-left-invariant.
\par
{
The continuum limit of $C_0(x,\mu)$ and the expression
of the order parameter in eq. (\ref{fop.2}) suggests that
in the region ${\cal C} \sim 0$ where confinement (or screening)
occurs there is condensation of vector meson pairs, while
in the deconfined region, where ${\cal C} \sim 1$, vector mesons
do not condense and are realized as particles.
}
%
%
%

%
%
\section{Note on Symmetry Breaking}
\label{sec:note}
The symmetry of the model and of the partition function
is rather interesting. The $SU(2)_L$ left transformations
(\ref{stck.5}) correspond to the local symmetries of the action, 
while the $SU(2)_R$ transformations (\ref{stck.5}) 
can only be  global symmetries, due to the fact that the mass
term in $S_E$ (eq. (\ref{lat.2})) breaks the  local
$SU(2)_R$ symmetry. For decreasing  mass parameter $m^2$ 
we expect the onset of a \underline{local}  $SU(2)_R$ symmetry. 
Then the fields $C(x,\mu)$, by construction (\ref{vec.1}), transform
according to a $SU(2)\otimes SU(2)\sim O(4) $ group of transformations
\begin{eqnarray}
C(x,\mu)' = g_R(x)C(x,\mu) g_R(x+\mu)^\dagger.
\label{vec.1p}
\end{eqnarray}
This fact has far reaching consequences. That is, in the limit of
zero mass, only closed loops of $C(x,\mu)$ fields have non zero
expectation value \cite{104915}. In particular all the correlators
in eqs. (\ref{vec.2.1}) are zero unless $y=x+\mu$ and $y+\nu=x$, i.e.
$\nu=-\mu$ and $C(y,\nu)= C(x,\mu)^\dagger$. But then the  $O(4)$ 
on the set $\{C_0(x),C_a(x)\}$ imposes
\begin{eqnarray}&&
C_{00,\mu\nu}(x,y)\simeq
C_{11,\mu\nu}(x,y)= C_{22,\mu\nu}(x,y) = C_{33,\mu\nu}(x,y) .
\label{vec.7}
\end{eqnarray}
The numerical simulations show that the onset of a local $SU(2)_R$
is very rapid when one crosses the line of PT. It becomes smooth for small
$\beta$, after the end point.
\par  
When the mass parameter becomes large the  $O(4)$ 
symmetry will be lost and only the global $SU(2)_R$ invariance will
survive and therefore  $C_{00,\mu\nu}(x,y)$ will
be substantially different from the $SU(2)_R$ - vector components.
%
%
%

%
%
\section{Survey}
\label{sec:survey}
We have performed standard Monte Carlo simulations
for the model based on the action (\ref{lat.2.2}). Heat bath 
has been used for the updating. Normally we have saved
a configuration every fifteen updatings for a total of 
10,000 measures.
\par
We considered  cubic 4-dimensional lattices with
periodic boundary conditions of different sizes:
$4^4,6^4,8^4,12^4,16^4,24^4$. We have chosen the size on
the basis of the precision needed. Typically a $6^4$
lattice size provides sensible results if $(\beta, m^2)$ 
is far away from the TL.
\par
\begin{figure}
\epsfxsize=100mm
\centerline{\epsffile{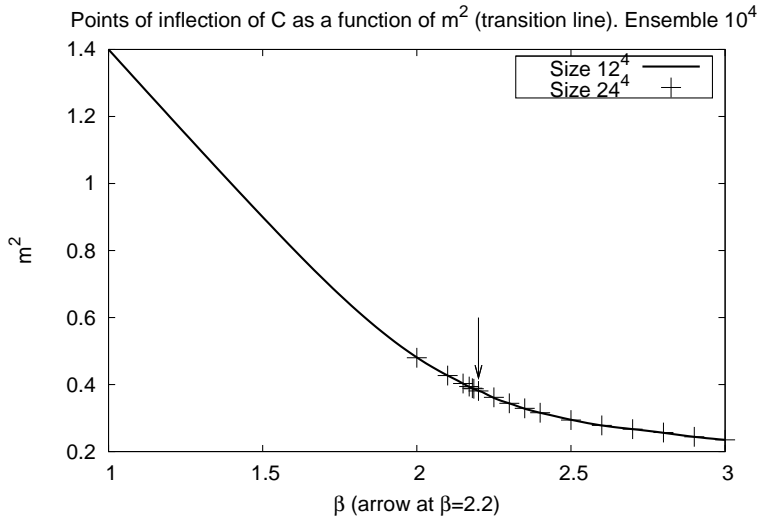}}
\caption{Phase diagram}
\label{fig.1}
\end{figure}
\begin{figure}[t]
\begin{center}
\includegraphics[clip,width=1.1\textwidth,height=0.5\textheight]
{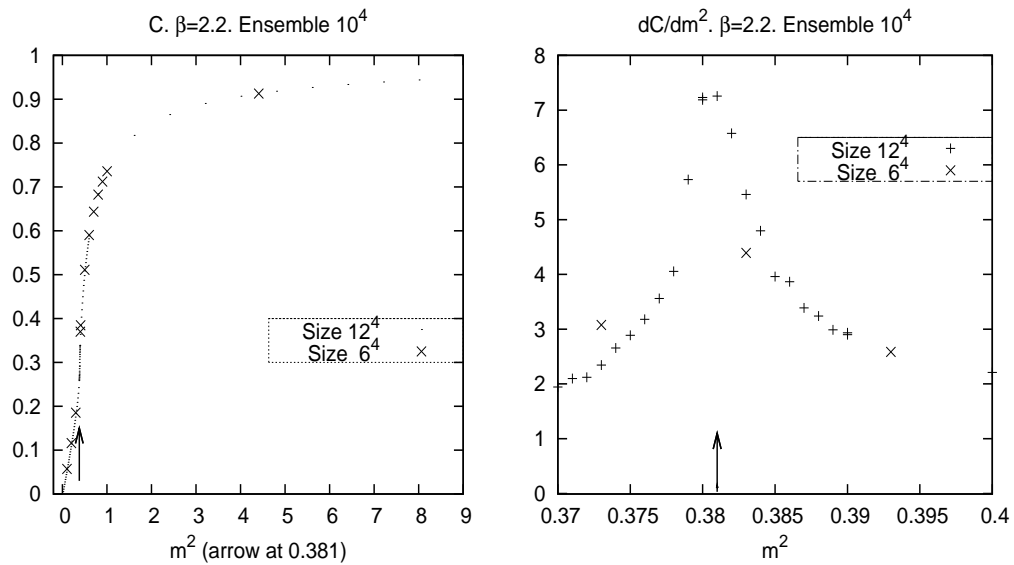}
\end{center}
\caption{${\frak C}$ and $\frac{\partial {\frak C}}{\partial m^2}$ 
at $\beta=2.2$ for size $6^4$ and $12^4$.}
\label{fig.2}
\end{figure}
\begin{figure}[t]
\begin{center}
\includegraphics[clip,width=1.1\textwidth,,height=0.5\textheight]
{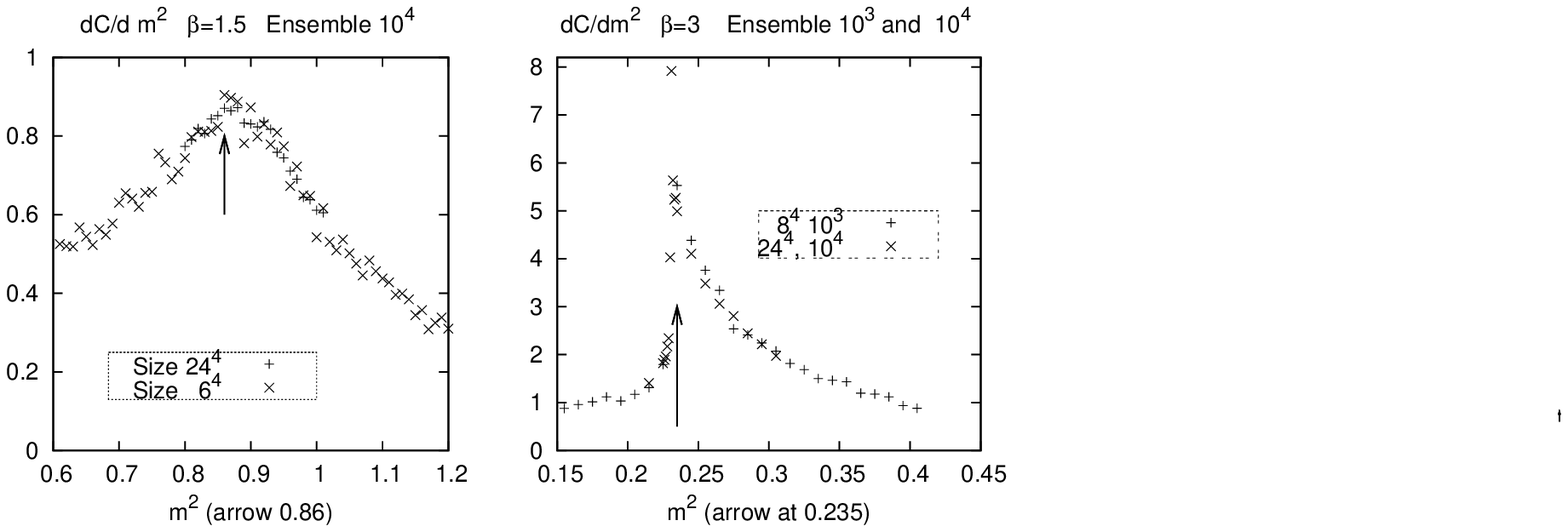}
\end{center}
\caption{$\frac{\partial {\frak C}}{\partial m^2}$ at $\beta=1.5$ and 
 $\beta=3$ for different lattice sizes.}
\label{fig.3}
\end{figure}

We have built a bird's-eye view of the region 
$\beta \in [1,4], m^2 \in [0,8]$
of some global quantities of the system. 
The quantities studied are those described in Section
\ref{sec:fop}, i.e. the energy per site $E$ (\ref{fop.1}), the order
parameter ${\mathfrak C}$ (\ref{fop.2}) and their derivatives with respect
to $m^2$ and $\beta$ as in eq. (\ref{vec.2.3})
\begin{eqnarray}&&
 \frac{\partial}{\partial m^2}{\mathfrak C}
= 
 \frac{\beta}{{D }}\Big\langle \Bigg(\frac{1}{\sqrt N}
\sum_{x\mu}C_0(x,\mu)  - \Big\langle\frac{1}{\sqrt N}
\sum_{x\mu}  C_0(x,\mu)\Big\rangle\Bigg)^2 \Big\rangle 
\nonumber\\&&
 \frac{\partial}{\partial m^2}E
= 
D(1+ \beta 
\frac{\partial}{\partial\beta}){\mathfrak C}
\nonumber\\&&
=  D \,\,\,{\mathfrak C} 
\nonumber\\&&
- \beta\Big\langle   \Big[\frac{1}{ \beta\sqrt N}S_E- \langle \frac{1}{ \beta\sqrt N}S_E\rangle\Big] 
\Big[\frac{1}{\sqrt N}\sum_{x\mu}  C_0(x,\mu)- \langle \frac{1}{\sqrt N}\sum_{x\mu}  C_0(x,\mu)\rangle\Big]\Big \rangle
\nonumber\\&&
 \frac{\partial}{\partial \beta} E
=  \Big\langle \Big[\frac{1}{ \beta\sqrt N}S_E- \langle \frac{1}{ \beta\sqrt N}S_E\rangle\Big]^2\Big \rangle
.
\label{survey.3}
\end{eqnarray}
Notice that
\begin{eqnarray}&&
 \frac{\partial}{\partial \beta}{\mathfrak C}
= 
 \frac{1}{{ND }} \frac{\partial}{\partial \beta}
 \Big\langle \Bigg(\sum_{x\mu}C_0(x,\mu)\Bigg)\Big\rangle =
 - \frac{1}{{D }} 
\nonumber\\&& \!\!\!\!\!
 \Big\langle \Bigg(\frac{1}{\sqrt N}\sum_{x\mu}C_0(x,\mu)-\langle \frac{1}{\sqrt N}\sum_{x\mu}  C_0(x,\mu)\rangle\Bigg)
 \Bigg(
 \frac{1}{ \beta\sqrt N}S_E- \langle \frac{1}{ \beta\sqrt N}S_E\rangle\Bigg)
 \Bigg\rangle.
\nonumber\\&&
\label{survey.4}
\end{eqnarray}
A survey on the parameters space shows a clear phase
change across the line represented in Fig. \ref{fig.1}.
In particular the line represents the {\sl loci} where
the dependence of the order parameter $\frak C$ from $m^2$ has a marked
inflection.
The line is stable under the change of  the size, for instance from $6^4$ 
through $24^4$.
%
A throughout study has shown that both energy and order
parameter are continuous everywhere including on the TL. 
Fig. \ref{fig.2} describes the dependence on  $m^2$ of $\mathfrak C$
 and its derivative.
All the first derivatives have a cusp behavior for $\beta \ge 2.2$ while
they are smooth for $\beta < 2.2$. Fig. \ref{fig.3}
exemplifies the situation for $\beta=1.5$ and $\beta = 3$.
There is some evidence for an end point at $\beta \sim 2.2, m^2\sim 0.381 $,
linked to the crossover point evidenced by early works on 
SU(2)-QCD simulations \cite{Creutz:1979dw}.
%
%
%
%
%

%
%
\section{Numerical Results}
\label{sec:num}
The numerical analysis of Sect. \ref{sec:survey} confirms 
the results obtained in previous works about the TL,
with some minor discrepancies, as the position
of the end point. On the {\sl vexata quaestio}, concerning
the order of the PT across the line and for $\beta \ge 2.2$,
our numerical evidence is not very conclusive, although we would be
more in favor of a second order type.
The present Section is devoted to the
new and surprising results. They show that the model is indeed
a faithful simulation of the massive Yang-Mills gauge theory
and moreover that unexpected and non trivial features
can be obtained in a region of the phase diagram unaccessible
by perturbation theory. 
\par
We study the operators (zero three-momentum)
\begin{eqnarray}
C_{a,\mu}(t):= \frac{1}{ \sqrt{N^\frac{3}{4}}}\sum_{\vec x}
C_a(\vec x, x_4,\mu)\biggr|_{x_4=t}, ~ a=0,1,2,3,~ \mu,\nu=1,2,3,4.
\label{num.1}
\end{eqnarray}
In particular we consider the two-point correlators
\begin{eqnarray}
C_{ab,\mu\nu}(t):=
\bigg\langle C_{a,\mu}(t+t_0) C_{b,\nu}(t_0)\bigg\rangle_C.
\label{num.2}
\end{eqnarray}
Numerical simulations support the selection rules
\begin{eqnarray}&&
C_{0b,\mu\nu}(t)= 0
\nonumber \\&&
C_{ab,\mu\nu}(t)\bigg |_{a\not = b}= 0, \quad a,b=1,2,3
\label{num.3}
\end{eqnarray}
imposed by the global $SU(2)_R$ invariance.
{
Moreover, according to eqs. (\ref{vec.1.1}), (\ref{vec.22.4})
and (\ref{vec.7}) we get  the limit values
\begin{eqnarray}
\lim_{m^2=0}C_{aa,\mu\nu}(0)=
0.25 \!\!!  \delta_{\mu\nu},\qquad a=0,1,2,3.
\label{num.3.11}
\end{eqnarray}
while for $t>0$ in the same limit the two-point functions are
expected to vanish according to the discussion of Section
\ref{sec:note}.
}
\par
\begin{figure}[t]
\begin{center}
\includegraphics[clip,width=1.1\textwidth,,height=0.4\textheight]
{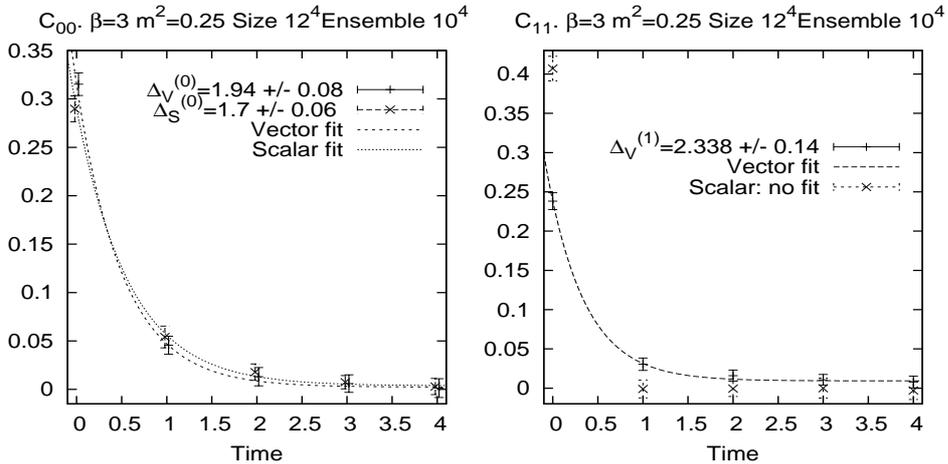}
\end{center}
\caption{{Correlators  at 
 $\beta=3$ and $m^2 = 0.25$: $C_{00}$ for the isoscalar part
and $C_{11}$ for the isovector according to eq. (\ref{num.3.1}).  
TL is at $m^2=0.235$.
The numerical values are the energy gaps for spin 1 ($\Delta_V^{(0)}$,
$\Delta_V^{(1)}$) and spin 0 ($\Delta_S^{(0)}$).}}
\label{fig.4}
\end{figure}
\begin{figure}[t]
\begin{center}
\includegraphics[clip,width=1.1\textwidth,,height=0.4\textheight]
{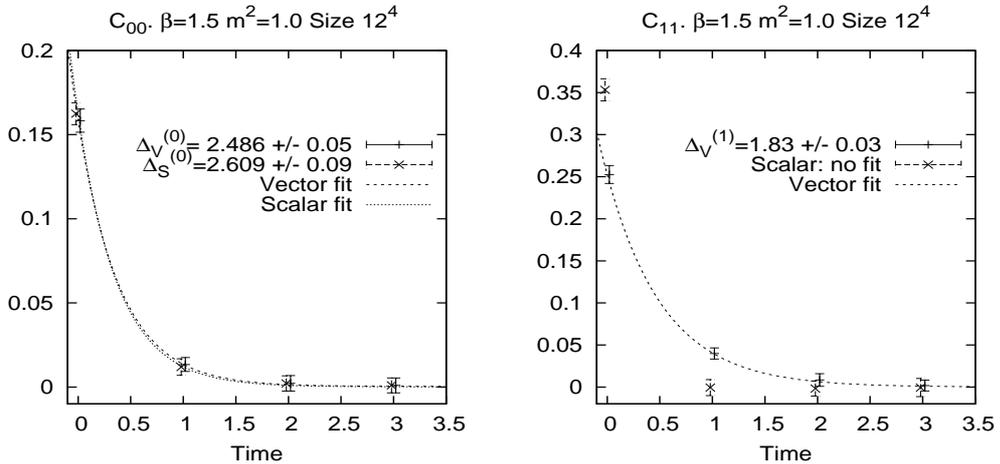}
\end{center}
\caption{{Correlators  at 
 $\beta=1.5$ and $m^2 = 1.0$ (TL is at $m^2=0.85$).
See caption of Fig.\ref{fig.4}.}}
\label{fig.5}
\end{figure}

The spin analysis is done by decomposing the correlators
into a spin one and spin zero parts (dots stand for { pairs 
of iso-indexes} $00$ or $11$,$22$,$33$ )
\begin{eqnarray}
C_{\cdots,\mu\nu}(t)= V_{\dots}(t)(\delta_{\mu\nu}- \delta_{4\mu}\delta_{4\nu})
+ S_{\dots}(t)\delta_{4\mu}\delta_{4\nu}.
\label{num.3.1}
\end{eqnarray}
We fit the amplitudes by a single exponential form
\begin{eqnarray}
F(t)= a + b e^{-t \Delta }.
\label{num.4}
\end{eqnarray}
A more complex analysis is not at reach with the data at hand. 
However the form turns out to be sufficient
for most of the cases that have been considered.
The energy gap $\Delta$ is obtained from a fit
on the correlator (\ref{num.2}) evaluated on $10^4$
configurations. Typical results are shown in Figs. \ref{fig.4}
and \ref{fig.5}.
\par
\begin{figure}[t]
\begin{center}
\includegraphics[clip,width=1.1\textwidth,,height=0.4\textheight]
{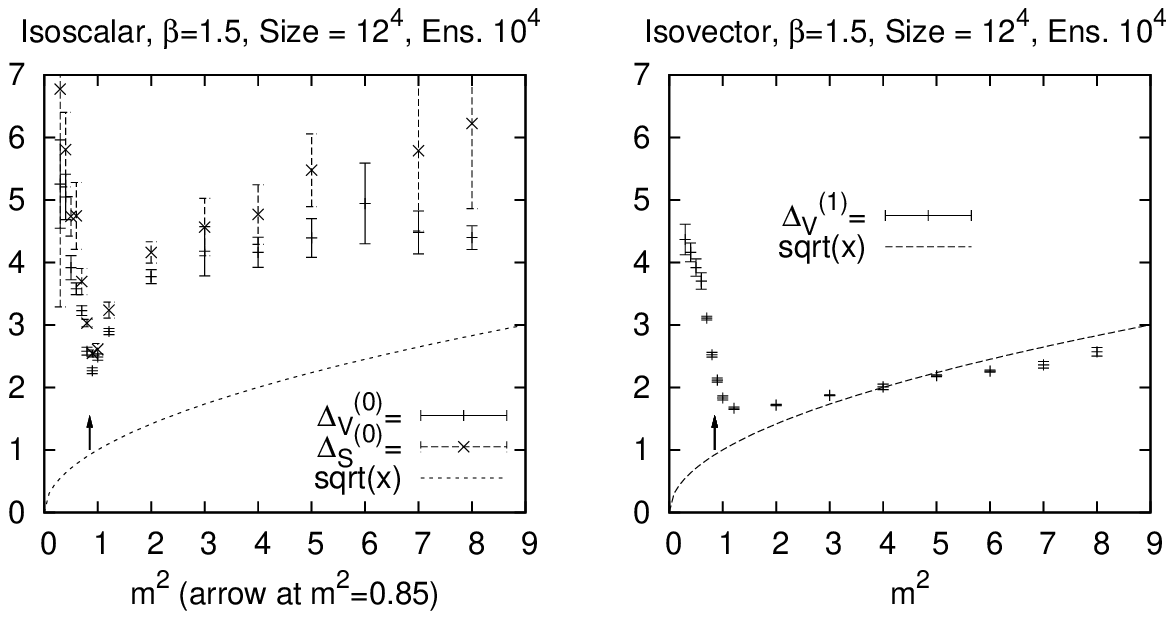}
\end{center}
\caption{On the y-axis the energy gaps of isoscalar spin one  
and spin zero (left)
and isovector spin one (right)
as function of $m^2$ for fixed  $\beta=1.5$
(TL at $m^2=0.85$).{The  energy gaps 
have been extracted
from fits as in Figs. \ref{fig.4} and \ref{fig.5} by using the expression
in eq. (\ref{num.4})}. The line is $\sqrt{|m^2|}$ and it is very close to
the spin 1 isovector energy gap in the region away from the 
phase transition.}
\label{fig.6}
\end{figure}
\begin{figure}[t]
\begin{center}
\includegraphics[clip,width=1.1\textwidth,,height=0.4\textheight]
{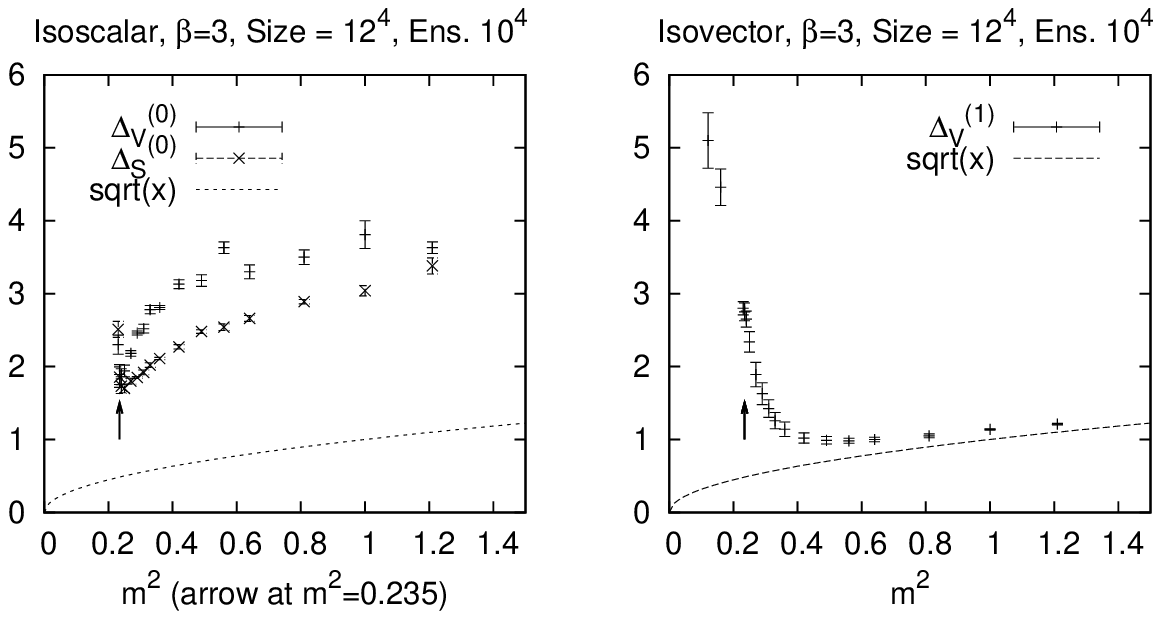}
\end{center}
\caption{Energy gaps of isoscalar spin one and spin zero (left)
and isovector spin one (right)
as function of $m^2$ for fixed  $\beta=3$
(TL at $m^2=0.235$).}
\label{fig.7}
\end{figure}

The energy gaps have been evaluated for several values of
$(\beta, m^2)$. Figs. \ref{fig.6} and \ref{fig.7}
represent the energy gaps as function of $m^2$. Several features 
are present in all the cases we have considered.
i) In the deconfined region 
and far from the TL the isovector correlator is due to a spin
one mode with energy gap $\Delta\simeq |m|$, while the isoscalar correlator
has both spin one and spin zero energy gaps, consistent with
a two-vector-meson threshold. ii) Near the TL the
isoscalar gaps become smaller than the threshold, thus suggesting
the presence of bound states. iii) Across the TL there
is a rapid increase of the gaps: within the errors all correlators
vanish for ${t}>0$ and a $O(4)$ symmetry is restored. The change of
phase is much more rapid for $\beta = 3$ than for $\beta=1.5$
(the change of scale of $m^2$ in Figs. \ref{fig.6} and \ref{fig.7}
should be properly accounted for).
%

%
%

\section{Conclusions}
We have investigated the deconfined phase of 
a massive Yang-Mills model by using a set of gauge
invariant fields. We give evidence of a 
TL in the parameters space $(\beta,m^2)$.  {
An order parameter $\cal C$ is introduced (the response to the
$m^2$ parameter) which is $\sim 1$ in the deconfined
region (large $m^2$), while $\sim 0$ below the TL 
(small $m^2$). The vanishing of the order parameter
corresponds to the condensation of pairs of vector
mesons. 
}
Far from this line the spectrum consists of an isovector
spin one meson and of  two-particle  states
in the isoscalar spin one and spin zero channels.
Moreover there is some evidence of bound
states near the TL in the isoscalar
channels for both spin states. 
 The presence of a discontinuity line confirms the
conjecture on {the existence of} two regimes: 
a low energy where the loop
expansion is valid and an extreme region where
massless Yang-Mills works. {The last point is a very
important step forward for the understanding 
of high energy processes.}

%
%

\section{Acknowledgements}
It is a pleasure to thank Bartolome All\'es for a 
stimulating introduction to the art of heat bath simulation 
and Davide Gamba
for invaluable help with the software during the early stage
of this work. 
\vfil\eject

\par\noindent
\end{document}